\documentclass[showpacs,twocolumn,aps,preprintnumbers,letterpaper]{revtex4}
\usepackage{amsmath,amssymb}
\usepackage{epsfig}
\usepackage{graphicx}
\usepackage{amsmath}
\usepackage{slashed}
\usepackage{amsfonts}
\usepackage{epstopdf}
\usepackage{changes}

\usepackage[pdftex, pdfborder= 0 0 0, citecolor=blue, urlcolor=blue, linkcolor=blue, colorlinks=true, bookmarksopen=true]{hyperref}

\newcommand{\be}{\begin{equation}}
\newcommand{\ee}{\end{equation}}
\newcommand{\bear}{\begin{eqnarray}}
\newcommand{\eear}{\end{eqnarray}}
\newcommand{\ba}{\begin{array}}
\newcommand{\ea}{\end{array}}

\def\be{\begin{eqnarray}}
\def\ee{\end{eqnarray}}
\def\bea{\be}
\def\eea{\ee}

\def\roughly#1{\mathrel{\raise.3ex\hbox{$#1$\kern-.75em%
\lower1ex\hbox{$\sim$}}}}

\definechangesauthor[name={Chang Hwan}, color=blue]{CHL}
\definechangesauthor[name={Hui Young}, color=orange]{RL}
\definechangesauthor[name={Ismail Zahed}, color=purple]{IZ}

\begin{document}

\title{Diffractive Vector Photoproduction using Holographic QCD}

\author{Chang Hwan Lee}
\email{clee@pusan.ac.kr}
\affiliation{Department of Physics, Pusan National University, Busan 609-735, South Korea }

\author{Hui-Young Ryu}
\email{huiyoung.ryu@pusan.ac.kr}
\affiliation{Department of Physics, Pusan National University, Busan 609-735, South Korea }

\author{Ismail Zahed}
\email{ismail.zahed@stonybrook.edu}
\affiliation{Department of Physics and Astronomy, Stony Brook University, Stony Brook, New York 11794-3800, USA}


\date{\today}
\begin{abstract}
We discuss diffractive photon-production of vector mesons in holographic QCD.  At large $\sqrt{s}$,
the QCD scattering amplitudes are reduced to the scattering of pair of dipoles exchanging 
a closed string or a pomeron.  We  use the 
holographic construction in AdS$_5$ to describe both the intrinsic dipole distribution in each hadron, 
and  the pomeron exchange. Our results for the  heavy meson   photon-production are made explicit
and compared to some existing experiments.
 \end{abstract}


\maketitle

\setcounter{footnote}{0}


\section{\label{sec:introduction}introduction}

Diffractive scattering at high energy is dominated by pomeron exchange, 
an effective object corresponding to the highest Regge trajectory. The slowly rising cross sections
are described by the soft Pomeron with a small intercep (0.08)  and vacuum quantum
numbers.  Reggeon exchanges have even smaller intercepts and are therefore subleading. Reggeon theory 
for hadron-hadron scattering with large rapidity intervals provide an effective explanation for the transverse 
growth of the cross sections~\cite{Gribov:1984tu}. In QCD at weak coupling the pomeron is described through
resummed BFKL ladders resulting in a large intercept and zero slope~\cite{Kuraev:1977fs,Balitsky:1978ic}.

The soft Pomeron kinematics suggests an altogether  non-perturbative approach. Through duality arguments, 
Veneziano suggested long ago that the soft Pomeron is a closed string exchange~\cite{Veneziano:1968yb}. In QCD
the closed string world-sheet can be thought as the surface spanned by planar gluon diagrams.
  The quantum theory of planar diagrams in supersymmetric gauge
theories is tractable in the double limit of a large number of colors $N_c$ and ${}^{\prime}$~t Hooft coupling
$\lambda=g^2N_c$ using the AdS/CFT holographic approach~\cite{HOLOXX}.

In the past decade there have been several attempts at describing the soft pomeron using
holographic QCD~\cite{Rho:1999jm,Janik:2000aj,Brower:2006ea,Stoffers:2012zw,Stoffers:2012ai,Basar:2012jb}.  
In this paper we follow the  work in~\cite{Stoffers:2012ai} and describe diffractive $\gamma+p\rightarrow V+p$ 
production through the exchange of a soft pomeron in curved AdS$_5$ geometry with a soft or hard wall. 
This is inherently a bottom-up approach~\cite{HOLOXXX} with the holographic or 5th direction
playing the role of the scale dimension for the closed string, interpolating between two fixed size dipoles. 
We follow the suggestion in~\cite{Polchinski:2001tt,Brodsky:2014yha} and describe the  intrinsic  dipole
size distribution of hadrons on the light cone through holographic wave functions in curved AdS$_5$. 
Diffractive production of vector mesons was investigated in the non-holographic context
by many in~\cite{MANY}. Recently a holographic description was explored in~\cite{Ahmady:2016ujw} in the context 
of the color glass condensate, and reggeized gravitons in~\cite{DJURIC}.

The organization of the paper is as follows:  In section 2 we briefly review the set up for diffractive scattering through
a holographic pomeron as a closed surface exchange in curved AdS$_5$ with a (hard) wall. In section 3, we detail the
construction of the light cone wavefunctions including their intrinsic light cone dipole distributions. In section 4  and 5 we make
explicit the AdS$_5$ model with a (soft) wall to descrive the intrinsic dipole distributions of massive vector mesons. As a
check on the intrinsic wavefunctions, we calculate the pertinent  vector electromagnetic decay constants. Our numerical
results for the partial cross sections and their comparison to vector photoproduction data are given in section 6. Our conclusions are summarized in section 7.

\section{\label{scattering} Dipole-Dipole Scattering}

\begin{figure}[!htb]
 \includegraphics[height=50mm]{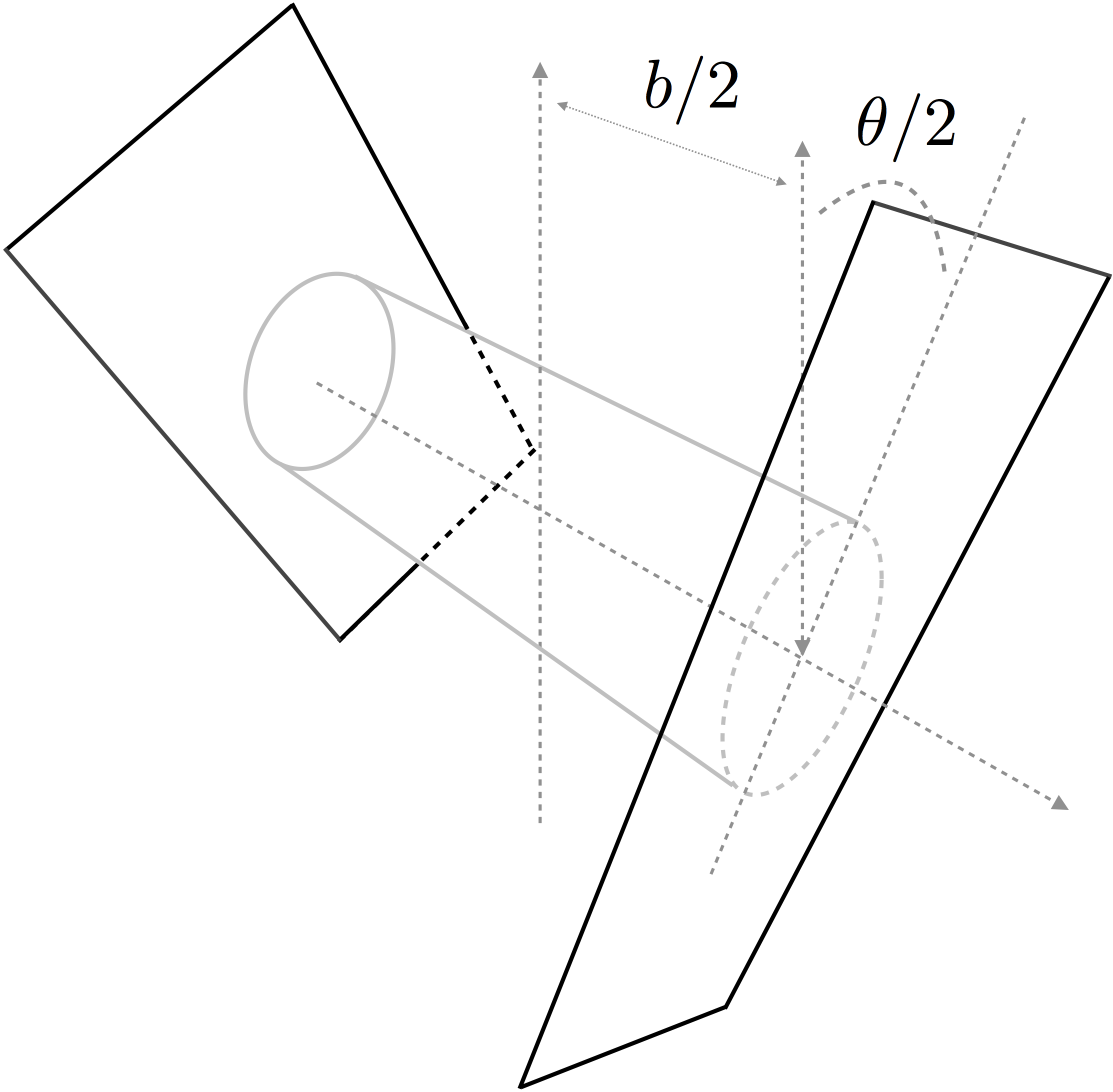}
  \caption{Dipole-Dipole Scattering.}\label{kinematics}
\end{figure}

In this section we briefly review the set-up for dipole-dipole scattering using an effective string theory.
For that we follow~\cite{Basar:2012jb} and consider the elastic scattering of two dipoles

\be
D_1 (p_1) + D_2 (p_2) \rightarrow D_1 (k_1) + D_2 (k_2)
\ee
as depicted in Fig.~\ref{kinematics}. 
$b$ is the impact parameter and the relative angle $\theta$ is the Euclidean analogue of the rapidity 
interval~\cite{Shuryak:2000df,Meggiolaro:1997mw}

\be
\cosh \chi = \frac{s}{2 m^2} -1 \rightarrow\cos \theta
\ee
with $s = (p_1 + p_2)^2$.

\subsection{Dipole-dipole correlator}

Following standard arguments as in~\cite{Basar:2012jb}, the scattering amplitude $\mathcal{T}$ in Euclidean space is given by

\bea
\frac{1}{- 2 i s}  \mathcal{T} (\theta, q) \approx 
\int d^2 \bold{b}~  e^{i \bold{q}_\perp \cdot \bold{b} } \,{\bf WW}
\eea
with ${\bf WW}$ the connected correlator of two Wilson loops, each represented by a rectangular loop 
sustained by a dipole and slated at a relative angle $\theta$  in Euclidean space as shown in Fig.~\ref{kinematics}. 
The leading $1/N_c$ contribution from a closed string exchange is

\be
\label{wwcalculation}
\bold{W} \bold{W}
 = g_s^2 \int \frac{d T}{2 T} \bold{K} (T)
\ee 
where
\be
\label{partitionfunction}
\bold{K} (T) = \int_{T}  \mathfrak{D}[x] ~ e^{- S[x] + {\rm ghost}}
\ee
is the string partition function on the cylinder topology with modulus $T$. 
The sum is over the string world-sheet with specific gauge fixing or ghost contribution. 
Here $g_s$ is the string coupling.

\subsection{Holographic Pomeron}

In flat $2+D_\perp$ dimensions, the effective string description for long strings
is the Polyakov-Luscher action with $D_\perp=2$. However, the dipole sources
for the incoming Wilson loops vary in size within a hadron. To account for this 
change and enforces conformality at short distances, we follow~\cite{Stoffers:2012zw}
and identify the dipole size $z$ with the holographic direction. The stringy exchange
in (\ref{wwcalculation}) is now in curved AdS in $2+D_\perp$ with $D_\perp=3$. 
At large  relative rapidity $\chi$ this exchange is dominated by the string tachyon 
mode with the result~\cite{Stoffers:2012zw}

\begin{eqnarray}
  \mathbf{WW}(z,z^\prime, {\bf b}_\perp) \simeq - \frac{g_s^2}{ 4}\frac{(2\pi^2)^{\frac 32}}{{\lambda}^{\frac 14}}
 \frac{(zz^\prime)^2}{z_0^4}
\mathbf{N} (\chi, z, z', \mathbf{b}_{\perp})     \label{eq:5}  
\end{eqnarray}
%
and 

\begin{eqnarray}
  \label{eq:7}
  \mathbf{N} (\chi, \mathbf{b}_{\perp} , z, z') 
= \frac{z_0^2}{zz'} \Delta(\chi, \xi)+
\frac{z}{z' } \Delta( \chi, \xi_* ),
\end{eqnarray}
$\Delta(\chi, \xi)$ refers to the tachyon propagator in  walled $AdS$.
It solves a curved diffusion equation  in the metric defined by

\be
ds^2=\frac{z_0^2}{z^2}\left(d{\bf b}_\perp^2+dz^2\right)
\ee
within $0\leq z\leq z_0$ with a zero current at the wall,

\begin{eqnarray}
  \label{eq:8}
  \Delta (\chi, \xi) = 
\frac{ \textrm{exp}\big[ (\alpha_{\textrm{ \bf P}}-1) \chi \big]
  }{ (4 \pi \textrm{\bf D} \chi)^{3/2} }
\frac{ \xi ~\textrm{exp}\big[ -\xi^2 /(4\textrm{\bf D}\chi)\big]  }{ \sinh (\xi)}
\end{eqnarray}
with the chordal distances given by

\begin{eqnarray}
  \label{eq:9}
  \cosh \xi &=& 
\cosh(u'-u) + \frac{ \mathbf{b}_{\perp}^2}{2z_0^2} e^{u'+u},  \nonumber \\
  \cosh \xi_* &=& \cosh(u'+u) + \frac{ \mathbf{b}_{\perp}^2}{2z_0^2}
                  e^{u'-u}.    \label{eq:10}
\end{eqnarray}
and $u={\rm ln}\frac {z_0}z$ and $u^\prime={\rm ln}\frac {z_0}{z^\prime}$.
The holographic Pomeron intercept and diffusion constant are respectively given by

\begin{eqnarray}
  \alpha_{\textrm{\bf P} } &=& 1+\frac{D_{\perp}}{12} -\frac{1}{2\sqrt{\lambda}},  \nonumber \\
\textrm{\bf D} &=&\frac{1}{2 \sqrt{\lambda}}. 
  \label{eq:11}
\end{eqnarray}
The string coupling in walled AdS is identified as
$g_s=\kappa_g\lambda/4\pi N_c$ and 
$\alpha^\prime/z_0^2=1/\sqrt{\lambda}$.
Here $\kappa_g$ is an overall dimensionless parameter that  takes into account the arbitrariness in the normalization
of the integration measure in (\ref{wwcalculation}). 
This analysis of the holographic Pomeron is different from the (distorted) spin-2 graviton exchange
 in~\cite{Brower:2006ea}
as the graviton is massive in walled AdS$_5$. Our approach is similar to the one followed in~\cite{Basar:2012jb} with the difference
that $2+D_\perp=5$ and not 10~\cite{Stoffers:2012zw}. It is an effective approach along the bottom-up scenario of AdS$_5$. 
Modulo different parameters, the holographic Pomeron yields a dipole-dipole total cross section that is similar to the one 
following  from  BFKL exchanges~\cite{Mueller:1994gb,Salam:1995zd}, and a wee-dipole density that is 
consistent with saturation at HERA~\cite{GolecBiernat:1998js}.

\section{Photon-hadron scattering}

In a valence quark picture an incoming meson is considered as a dipole made of a $q\bar q$ pair, while a baryon
is considered as a dipole made of a  pair of a quark-diquark. The quantum scattering amplitude follows
by assigning to the scattering pairs dipole sizes $r_{1,2}$ and distributing them within the quantum mechanical amplitude of the 
pertinent hadron. At large $\sqrt{s}$ the scattering particles propagate along the light cone and are conveniently described by
light cone wave functions. Typically, the latters are given in terms of an intrinsic wavefunction  $\Psi(x,r)$ for a dipole of size
$r$ with a fraction of parton longitudinal momentum $x$. 
With this in mind, the scattering amplitude for the diffractive process for vector meson photo-production  $\gamma+p\rightarrow V+p$, reads

\begin{eqnarray}
\label{AGP}
{\cal A}=&&-2is\int d^2{\bf b}_\perp e^{-iq_\perp b_\perp}\\
&&\times\int \frac{d^2r_1dx_1}{4\pi}\Psi_V^\dagger\Psi_\gamma(x_1,r_1)\nonumber\\
&&\times\int \frac{d^2r_2dx_2}{4\pi}\Psi_p^\dagger\Psi_p(x_2,r_2)\, (-{\bf WW}(r_1,r_2,{\bf b}_\perp))\nonumber\\
=&&-2is\int d{\bf b}^2_\perp \pi\,J_0\left(|{\bf b}_\perp |\sqrt{|t|}\right)\nonumber\\
&&\times\int \frac{d^2r_1dx_1}{4\pi}\Psi_V^\dagger\Psi_\gamma(x_1,r_1)\nonumber\\
&&\times\int \frac{d^2r_2dx_2}{4\pi}\Psi_p^\dagger\Psi_p(x_2,r_2)\, (-{\bf WW}(r_1,r_2,|{\bf b}_\perp|))\nonumber\\
\end{eqnarray}
The $\frac 1{4\pi}$ normalization conforms with the light cone rules.

Note that  in flat $D_\perp$-space (also for  $\xi\ll 1$), the propagator (\ref{eq:8})  simplifies 

\bea
\label{flat}
\Delta_F(\chi, \xi)=&&\Delta_F(\chi, \xi_*)\nonumber\\
=&&
\frac{ \textrm{exp}\big[ (\alpha_{\textrm{ \bf P}}-1) \chi \big]}{ (4 \pi \textrm{\bf D} \chi)^{3/2} }\nonumber\\
&&\times\textrm{exp}\big[ -({\bf b}_\perp^2+(z-z^\prime)^2) /(2\alpha^\prime \chi)\big]  \nonumber\\
\eea
after the substitution  $z_0^2{\bf D}\rightarrow \frac {\alpha^\prime}2$. For  an estimate of (\ref{AGP}) we may insert (\ref{flat}) into (\ref{AGP}), ignore the wall and assume $z\sim z^\prime$ to carry out the integration in (\ref{AGP}) exactly 

\begin{eqnarray}
\label{AGPFLAT}
{\cal A}_F=&& 2\times \frac{g_s^2}{ 4}\frac{(2\pi^2)^{\frac 32}}{{\lambda}^{\frac 14}}\frac{2i sz_0^2}{(4\pi {\bf D}{\rm ln}\frac s{s_0})^{\frac 12}}\,\left(\frac{s}{s_0}\right)^{\alpha_{\bf P}(t)-1}\nonumber\\
&&\times\int \frac{d^2r_1dx_1}{4\pi}\Psi_V^\dagger\Psi_\gamma(x_1,r_1)\nonumber\\
&&\times\int \frac{d^2r_2dx_2}{4\pi}\Psi_p^\dagger\Psi_p(x_2,r_2)\, 
\end{eqnarray}
with the Pomeron trajectory 

\be
\alpha_{\bf P}(t)=\alpha_{\bf P}+\frac{\alpha^\prime}2t
\ee

\subsection{Photon wave function}

The description of the light cone photon wave function in terms of a $q\bar q$ pair follows from 
light cone perturbation theory as described in~\cite{Lepage:1980fj}. Let $Q^2$ be the virtuality of the photon of polarization $h$.
The amplitude for finding a  $q\bar q$ pair in the virtual photon with light cone momentum fractions $(x,\bar x)$ is given by
\cite{MANY,Lepage:1980fj}

\begin{eqnarray}
  \label{eq:12X0}
 && \psi_{h, \bar h}^{\gamma, L}  (r,x:Q^2, m_f) 
=\sqrt{N_c}\,e e_f \,
\delta_{h,-\bar h}
 x\bar x\, 2Q \frac{K_0 (\epsilon r)}{2\pi}\nonumber \\
  &&\psi_{h, \bar h}^{\gamma, T\pm}  (r,x:Q^2, m_f)=
\sqrt{2N_c}\,e e_f \,
\Big[
 i e^{\pm i\theta_r}( x \delta_{h\pm}\delta_{ \bar h \mp} \nonumber\\
 &&-\bar x
\delta_{h\mp}\delta_{ \bar h\pm}) (\mp \partial_r  )
+m_f \delta_{h\pm}\delta_{ \bar h\pm})
\Big] \frac{K_0(\epsilon r)}{2 \pi}
\end{eqnarray}
with $\Psi^\gamma_{h, \bar h}$ the matrix entries in helicity of $\Psi_\gamma$ in (\ref{AGP}).
Here $ee_f$ is the charge of a quark of flavor $f$,
$\epsilon^2 = x\bar xQ^2 + m_f^2$, and $K_{0,1}$ are  modified Bessel functions. 
Also $(r,\theta_r)$  are the 2-dimensional dipole polar coordinates. While the photo-production 
analysis  to be detailed below corresponds to $Q^2=0$, we will carry the analysis for general $Q^2$ for future reference.

\subsection{Hadron wave functions}

We start by defining the proton (squared) wave function for a pair of quark-diquark as 

\begin{eqnarray}
\left|\psi_p(x, r)\right|^2=\frac 2{r_p}\delta\left(x-\frac 12\right)\delta(r-r_p)
\end{eqnarray}
by simply assuming equal sharing of the longitudinal momentum among the pair,
and a fixed dipole size $r_p$, with the normalization

\begin{eqnarray}
\int \frac{d^2r\, dx}{4\pi}\left|\psi_p(x, r)\right|^2=1
\end{eqnarray}

The vector meson wave function on the light cone will be sought by analogy with the
photon wave function given above. Specifically we write


\begin{eqnarray}
  \label{eq:12X1}
 && \psi_{h, \bar h}^{V, L}  (r,x:M_V, m_f) 
=\delta_{h,-\bar h}\, x\bar x\, f_L(x,r)\nonumber\\
&&  \psi_{h, \bar h}^{V, T\pm}  (r,x:M_V, m_f) =
\bigg[
\frac{i r^{\pm}}{rM_V}
(x \delta_{h\pm }\delta_{\bar h \mp} 
\nonumber \\
&{}&
- \bar x \delta_{h\mp }\delta_{\bar h \pm})(\mp \partial_r)
+\frac{m_f}{M_V} \delta_{h \pm}\delta_{\bar h \pm}
\bigg] \,f_T(x,r)
  \label{eq:13}
\end{eqnarray}
where $\Psi^V_{h, \bar h}$ are the matrix entries in helicity of $\Psi_V$ in (\ref{AGP}).
The intrinsic $f_{L,T} (x,r)$ dipole distributions for the vector mesons will be sought below 
in  the holographic construction by identifying the holographic direction in the description
of massive vector mesons with the dipole size~\cite{Polchinski:2001tt,Brodsky:2014yha}.

\subsection{Partial cross sections}

The partial diffractive cross sections for the production of longitudinal and transverse
vector mesons are given by

\begin{eqnarray}
\frac {d\sigma_{L,T}}{dt}=\frac{1}{16\pi s^2}\left|{\cal A}_{L,T}\right|^2
\end{eqnarray}
with the virtual-photon-vector-meson transition amplitudes following from the contraction of the
helicity matrix elements (\ref{eq:12X0}-\ref{eq:12X1}).  The results are

\begin{eqnarray}
L:\, &&\Psi^\dagger_V \Psi_\gamma=\frac{2\sqrt{N_c}}{\pi}\,ee_V\,(x\bar x)^2\,Q{K_0}\,{f_L(x,r)}\nonumber\\
T:\,&&\Psi^\dagger_V \Psi_\gamma=\frac{\sqrt{N_c}}{\pi\sqrt{2}}\,ee_V\,\,\nonumber\\
&&\times\left(\frac{\epsilon}{M_V}(x^2+{\bar x}^2)\,K_1\,(-\partial_r)\,+\frac{m_f^2}{M_V}\,K_0\right)
\,{f_T(x,r)}\nonumber\\
\label{GV}
\end{eqnarray}
The vector charge $e_V$ is  computed as the average charge 

\be
\label{EV}
e_V=\left|\sum_f a_f e_f\right|=\left(\rho: \frac 1{\sqrt{2}}, \omega: \frac 1{\sqrt{18}}, \phi: \frac 13, J: \frac 23, \Upsilon: \frac 13\right)
\nonumber\\
\ee
in a state with flavor content $V=\sum_fa_f\bar f f$. 
The elastic differential cross section follows as

\be
\frac{d\sigma_{\rm el}}{dt}=\frac {d\sigma_{L}}{dt}+\frac {d\sigma_{T}}{dt}
\ee

\section{$f_{L,T}$ from holography}

The intrinsic light cone distributions in the vector mesons is inherently non-perturbative. 
Our holographic set-up for the description of the $\gamma+p\rightarrow V+p$ process 
as a dipole-dipole scattering through a holographic pomeron in AdS$_5$ suggests that
we identify  the intrinsic light cone distributions $f_{L,T}$ with the holographic wave function
of massive Spin-1 mesons in AdS$_5$. The mass will be set through a tachyon field in bulk.

\subsection{AdS model for Spin-1}

With this in mind, 
consider an AdS$_5$  geometry  with a vector gauge field ${A}$ and a dimensionless 
tachyon field ${X}$  described by the non-anomalous action

\begin{eqnarray}
&&S=\int d^4xdz\nonumber\\
&& \left(\frac 1{2g_5^2} \frac 1z F^{MN}F_{MN}-\frac 1{z^3}\left|DX\right|^2+\frac 3{z^5}\left|X^2\right|\right)
\end{eqnarray}
with $DX=dX+{A}{X}$ and ${F}=d{ A}$,  $M,N=0,1,2,3,z$ and signature $(-, +,+,+,+)$
The coupling $g^2_5\equiv 12\pi^2/N_c$ is fixed by standard arguments~\cite{HOLOXXX}. 
The background tachyon  field satisfies

\be
\label{BB3}
\frac{d}{dz}\left(\frac{1}{z^3}\frac{dX}{dz}\right)+\frac{3}{z^5}X=0
\ee
which is solved by

\be
\label{BB4}
X(z)\approx c_1z+c_2z^3
\ee
The constants in (\ref{BB4}) are fixed by the holographic dictionary~\cite{HOLOXX,HOLOXXX} near the UV boundary
($z\approx 0$)

\be
\label{BB5}
X(z)\approx Mz+\left<\bar QQ\right>z^3
\ee
In the heavy quark limit $\left<\bar Q Q\right> \rightarrow 0$,  so $X(z)\approx Mz$.


In the presence of $X(z)$, the vector gauge field satisfies 

\begin{eqnarray}
D^MF_{MN}+\frac {4g_5^2}{z^2}X^2 A_N=0
\end{eqnarray}
We now seek a plane-wave vector meson with 4-dimensional spatial polarization $\epsilon_\mu$ in the form

\begin{eqnarray}
A_M(x,z)=e^{ix^\mu p_\mu}\left(\sqrt{z}\varphi(z)\right)\,\delta_{M\mu}\,\epsilon_\mu(p)
\end{eqnarray}
which yields

\begin{eqnarray}
-\varphi^{\prime\prime}+\frac {\frac 34}{z^2}\varphi+\frac {4g_5^2}{z^2}X^2(z)\varphi=-p^2\varphi
\end{eqnarray}
We now use the solution for $X(z)\approx c_1z+c_2 z^3$ with $c_2=0$  (no heavy chiral condensate),
 and identify $4g_5^2c_1^2=(2m_f)^2$ with $m_f$ the (constituent) quark mass. Thus near the
boundary

\begin{eqnarray}
\label{00}
-\varphi^{\prime\prime}+\frac {\frac 34}{z^2}\varphi\approx (-p^2-(2m_f)^2)\varphi
\end{eqnarray}
We can now either solve (\ref{00}) using a hard-wall by restricting (\ref{00}) to the slab geometry
$0\leq z\leq z_0$, or introducing a soft wall~\cite{SOFT}. The former is a Bessel function with a spectrum that
does not Reggeizes, while the latter is usually the
one favored by the light-cone with a spectrum that Reggeizes.  The minimal soft wall amounts

\begin{eqnarray}
\label{01}
-\varphi^{\prime\prime}+\frac {\frac 34}{z^2}\varphi +\kappa^2z^2\varphi=(-p^2-(2m_f)^2)\varphi
\end{eqnarray}
Defining $E=M^2-(2m_f)^2$, it follows that

\begin{eqnarray}
\label{001}
M_n^2=&&4\kappa^2(n+1)+(2m_f)^2\nonumber\\
\varphi_n(z)\sim &&({\kappa}z)^{\frac 32}\,e^{-\frac 12{\kappa^2z^2}}\,L_n^1(\kappa^2z^2)
\end{eqnarray}
The meson spectrum Reggeizes. The 
value for $\kappa=\sqrt{\sigma_T}/2\approx \frac 12 {\rm GeV}$ is fixed by the string tension. 


\subsection{Intrinsic wave functions}
We now suggest that the holographic wavefunction 
\begin{eqnarray}
\label{02}
\varphi_{n=0}(z)\sim ({\kappa}z)^{\frac 32}\, e^{-\frac 12{\kappa^2z^2}}
\end{eqnarray}
can be related to the intrinsic amplitudes $f_{L,T}$ for the dipole distribution 
in the light cone wavefunctions for the vector mesons in (\ref{eq:12X1}). For that we note that the main
part of the transverse vector in (\ref{eq:12X1}) 
satisfies  $\vec\Psi^{T}\sim \vec \nabla f_T$.  With this in mind, 
we identify the holographic coordinate $z$ with the relative dipole size $r$ through
$z=\sqrt{x\bar x}r$~\cite{Polchinski:2001tt,Brodsky:2014yha}, and match
the r-probability of the intrinsic state to the z-probability of
the spin-1 state in bulk AdS$_5$,

\be
 dr\,\left|\sqrt{r}\,\vec\nabla f_T\right|^2\sim 
\left( \frac{dz}{z}\left|\sqrt{z}\varphi_0(z)\right|^2\right)_{z=\sqrt{x\bar x}r}
\ee
 The extra $\frac 1z$ in the bracket is the warping factor. Solving for  $f_T$ we obtain 

\be
\label{FTXR}
f_T(x,r) =2\kappa (x\bar x)^{\frac 12}e^{-\frac 12\kappa^2x\bar x {r^2}}
\ee
which normalizes to 1

\begin{eqnarray}
\int \frac {d^2r dx}{4\pi}\left|f_T(x,r)\right|^2=1
\end{eqnarray}

For  a massive spin-1 meson with the helicity content and quark mass analogous to the $\gamma^*\sim \bar qq$ 
content as ansatz  in (\ref{FTXR}), we will assume the holographic  dipole content  derived in  (\ref{FTXR}),
with instead general  overall  constants

\begin{eqnarray}
\label{OURS}
f_{T,L}(x,r)\rightarrow N_{T,L} \, (x \bar x )^{\frac 12}\, e^{-\frac 12 {\kappa^2 x\bar xr^2}}
\end{eqnarray}
 $N_{T,L}$ are now fixed by the helicity-dependent normalizations using (\ref{eq:12X1}), i.e. 

\begin{eqnarray}
\int \frac {d^2r dx}{4\pi}\left|\psi^{L,T}_V(x,r)\right|^2=1
\end{eqnarray}
More specifically we have

\begin{eqnarray}
&&L:\, \frac{N_L^2}{2\kappa^2}\int_0^1(x\bar x)^2 \,dx=1\nonumber\\
&&T:\, \frac{N_T^2}{2\kappa^2}
\int_0^1\left(\frac{m_f^2}{M_V^2}
+x\bar x(x^2+{\bar x}^2)
\frac{\kappa^2}{2M_V^2}  \right)dx=1
\label{NORM}
\end{eqnarray}
which fix $N_{T,L}$

\begin{eqnarray}
&&L:\,  N_L=\kappa\sqrt{60}\nonumber\\
&&T:\,  N_T=
\kappa M_V\left(\frac {40}{\kappa^2+20m_f^2}\right)^{\frac 12} 
\label{NORMX}
\end{eqnarray}
(\ref{OURS}) is in agreement with the intrinsic dipole wave function developed in~\cite{Brodsky:2014yha} 
using the light cone holographic procedure for $m_f=0$. We note that  (\ref{02})
 describes a massive spin-1 gauge field in AdS$_5$.

\section{Leptonic Decay constants}

The size of the  light cone wavefunction  is  empirically constrained by 
the electromagnetic decay width $V\rightarrow e^+e^-$ as captured by the 
measured vector decay constant $f_V$ for each of the vector mesons,

\begin{eqnarray}
\label{MATRIX}
\left<0\biggl|J_{\rm em}^\mu(0)\biggr|\psi_V^{L,T}(q)\right>=ef_VM_V\epsilon_{L,T}^\mu(q)
\end{eqnarray}
This  puts an empirical constraint on the longitudinal 
and transverse light cone wavefunctions  (\ref{FTXR}) using the holographic intrinsic wavefunctions (\ref{OURS}) as 
suggested earlier.

\subsection{Longitudinal}

More specifically, the longitudinal wavefunction gives for the right-hand-side in (\ref{MATRIX})

\begin{eqnarray}
ef_VM_V\epsilon^+_L (q)
\rightarrow ef_Vq^+
\end{eqnarray}
as $q^+\rightarrow \infty$, with the conventions $q^2=q^+q^-=-Q^2=-M_V^2$.  
The left-hand-side in (\ref{MATRIX}) can be reduced using the light cone  rules in the Appendix 
of~\cite{Lepage:1980fj} together with the  longitudinal wavefunction (\ref{eq:12X1}) to have

\begin{eqnarray}
&{}&\left<0|J_{\rm em}^+(0)|\psi_V^{L}(q)\right>=\int_0^1dx
\left(ee_V\sqrt{N_c}\,\delta_{h,-\bar h}\,\frac{2\sqrt{x\bar x}}{{\sqrt{x}}\sqrt{\bar x}} q^+\right)\nonumber\\
&{}&\qquad
\times\left(\int \frac {d^2r d^2k}{(2\pi)^3}e^{ik\cdot r}\delta_{h,-\bar h}x\bar x\, f_L(x,r)\right)
\end{eqnarray}
The first bracket refers to the reduction of the current, and the second bracket to the
reduction of the longitudinal wavefunction. 
 The result for the 
vector decay constant from the longitudinal current $J_{\rm em}^+$ is
\be
\label{FVL}
\frac{f^L_V}{\kappa}=e_V\sqrt{N_c}\,\frac{3\sqrt{15}}{32}
\ee
after the use of the normalization $N_L$ as given in (\ref{NORMX}). For example, for the rho meson 
 $f_\rho/\kappa=9\sqrt{5}/(32\sqrt{2})$, 
while for the phi meson  $f_\phi/\kappa=3\sqrt{5}/32$.

\subsection{Transverse}

For a consistency check, the same rules apply to the transverse component of the current $J_{\rm em}^{1}$.
The transverse wavefunction gives for the right-hand side of (\ref{MATRIX})

\be
\label{TT1}
ef^T_VM_V\epsilon_T^1(q)\rightarrow ef^T_VM_V\left(\frac{-1}{\sqrt{2}}\right)
\ee
The left-hand-side can be reduced using also the light cone rules

\bea
\label{TT2}
&&\left<0|J_{\rm em}^1(0)|\psi_V^T(q)\right>=\int_0^1\,dx\int \frac{d^2rd^2k}{(2\pi)^3}e^{ik\cdot r}\nonumber\\
&&\times\frac {ee_V\sqrt{N_c}}{\sqrt{2x2\bar x}}
\left(\frac{xk^--\bar x k^+}{\sqrt{x\bar x}}\delta_{h,-\bar h}
-
 \frac{m_f}{\sqrt{x\bar x}}\delta_{h,\bar h}\right) \nonumber\\
&&\times
\left(\frac {i r^\pm}{M_V}
\left( - \frac
  1r\partial_r\right)(x\delta_{h+}\delta_{\bar h -}-\bar
x\delta_{h-}\delta_{\bar h+})+\frac
{m_f}{M_V}\delta_{h,\bar h}\right)
\nonumber \\
&{}&\times\,f_T(x,r)
\eea
The first contribution stems from the reduction of the current and the second contribution from the reduction of the
transverse wavefunction. The $\mp$ signs in (\ref{TT2}) follows the $h=\pm$ assignments.  Using the explicit
form of the wavefunction (\ref{FTXR}) and performing an integration by parts, we have the identity 
\be
\label{TT3}
\int \frac{d^2rd^2k}{(2\pi)^3}e^{ik\cdot r}\,ix^+k^-\left(-\frac
  1r\partial_r\right)\,f_T(x,r)=-
\frac{\kappa^2 x\bar x}{\pi} \,f_T(x,0)
\nonumber\\
\ee
Inserting (\ref{TT3}) in (\ref{TT2}) gives for the left-hand-side
\bea
\label{TT4}
&&\left<0|J_{\rm em}^1(0)|\psi_V^T(q)\right>
=-\frac{ee_V\sqrt{N_c}}{2\pi}\nonumber\\
&&\int_0^1dx\left(\frac{m_f^2+\kappa^2 x\bar x(x^2+{\bar x}^2)}{M_V x\bar x}\right)\,f_T(x,0)
\eea
which reduces to
\be
\label{FVT}
\frac{f^T_V}{\kappa}=&&\frac{e_V\sqrt{N_c}}{\pi\sqrt{2}}\,
\left(
\frac {40M_V^2}{\kappa^2+20m_f^2} 
\right)^{\frac 12}\nonumber\\
&&\times\int_0^1\frac{dx}{\sqrt{x\bar
    x}}\,\left(\frac{m_f^2}{M_V^2}+x\bar x(x^2+{\bar
    x}^2)\frac{\kappa_T^2}{M_V^2}\right)
\ee

\begin{figure}[t]
  \begin{center}
  \includegraphics[width=8cm]{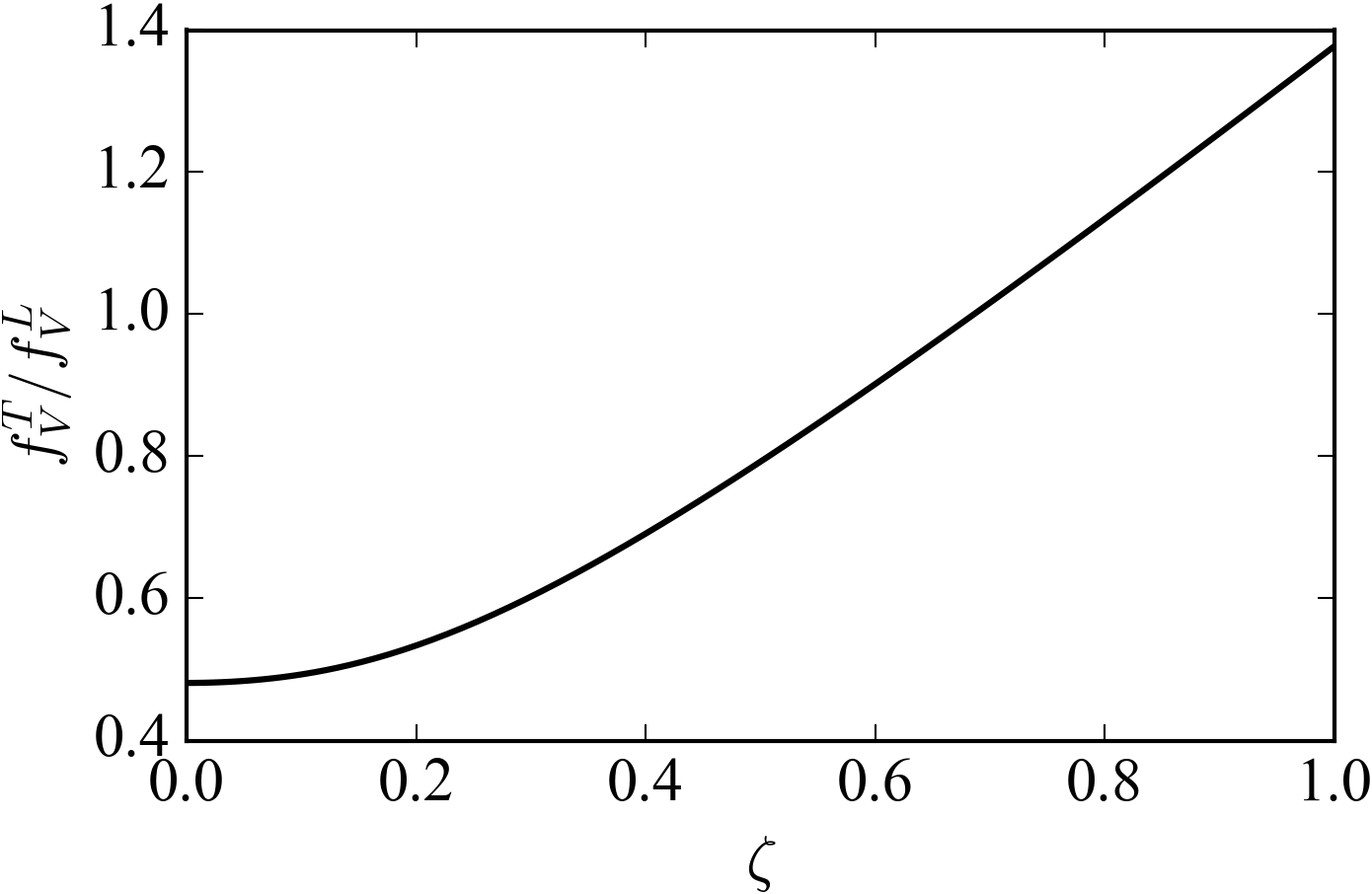}
  \caption{Ratio (\ref{RATIOX})  of the transverse to longitudinal decay constants
  vs $\zeta=2m_f/M_V$ from the massless to the heavy quark limit.}
    \label{fig_ratio}
  \end{center}
\end{figure}

Substituting the value of $\kappa$ from the Regge spectrum (\ref{001}) yields the transverse to longitudinal ratio
for the decay constants

\be
\label{RATIOX}
\frac{f_V^T}{f_V^L}=
\frac{1}{6 \sqrt{3}} \frac{59 \zeta^2 +5}{(1+19 \zeta^2)^{\frac{1}{2}}}
\ee
with $\zeta=2m_f/M_V$. 
In Fig.~\ref{fig_ratio} we show the behavior of (\ref{RATIOX}) in the range $\zeta=0,1$ 
from the massless to the heavy quark limit where it reaches 1.

%
\begin{table}[ht]
\caption{Holographic parameters along with the model prediction for 
$M_V,f_V$. See text.}
\label{table:1}
\begin{tabular}{c|@{\hskip .13in}c@{\hskip .1in}c@{\hskip
  .12in}c@{\hskip .12in}c@{\hskip .12in}c@{\hskip .1in} c @{\hskip .1in} } 
\hline\hline
&$M_V$&$f^L_V$&$f_V^T/f_V^L$ &$m_f$& $\kappa$  &  $g_s$ \\
& $(M_V)_{\textrm{exp}}$ &$(f_V) _{\textrm{exp}}$ &&&   &\\
&[GeV]&[MeV] &[MeV] &[GeV] && \\
\hline
$\rho$       & 1000 & 186.9 &1.087 &$0.380$& 0.325 &   0.63 \\
               &(775.3) &(204) & && &\\
$\omega$ & 971  & 44.8 &  1.114&$0.380$& 0.302 &  1.46\\
               & (782.7)&(59) &    &    && \\
$\phi$      & 1172    &78.67 & 1.096 &$0.450$&0.375 & 0.44\\
             & (1020)  &  (74) &   &    & &\\
$J/\Psi$    & 3185   &153.3 & 1.344 &$1.550$&0.366 &  1.50 \\
              & (3097) & (90)  &   &    & &\\
$\Upsilon$&  9472 & 49.0 & 1.367 &4.730 &0.234 &   5.00 \\
             &(9460)  &  (25) &   &    & &\\
\hline
   & $s_0$  &  0.1 GeV${}^2$ &  &  & $D_\perp$  &3    \\
   & $z_p$  &1.8 GeV${}^{-1}$&   & & $\lambda$ &23  \\
\hline
\hline
\end{tabular}
\end{table}

\section{Numerical analysis}

To carry out the numerical analysis, we can 
 partially eliminate the model dependence in the tansition amplitudes (\ref{GV}), by  trading
$\kappa$ in the  normalizations $N_{L,T}$ in (\ref{NORMX}) with (\ref{FVL}) to obtain




\bea
L:\, &&\Psi^\dagger_V \Psi_\gamma=\frac{128}{3\pi}ef_V^L
(x\bar x)^2\,Q{K_0}\,{\mathbb F (x,{\kappa}r)}\nonumber\\
T:\,&&\Psi^\dagger_V \Psi_\gamma=\frac{128}{3\pi}ef_V^L
\bigg(\frac 1{3+57\zeta^2}\bigg)^{\frac 12}\nonumber\\
&&\times\left(\frac{\epsilon}{M_V}(x^2+{\bar x}^2)\,K_1\,(-\partial_r)\,+\frac{m_f^2}{M_V}\,K_0\right)
\,{\mathbb F (x,{\kappa}r)}\nonumber\\
\label{GVVX}
\eea 
Here we have set 

\begin{eqnarray}
\label{NEW}
\mathbb F (x,\kappa r)= (x \bar x )^{\frac 12}\, e^{-\frac 12 {\kappa^2 x\bar xr^2}}
\end{eqnarray}
with $\kappa$  fixed by the ground state meson mass in  (\ref{001}) 

\be
\kappa=\frac{M_V}2\sqrt{1-\zeta^2}
\label{KAPPA}
\ee

With the exception of $g_s, \kappa, m_f$, all holographic  parameters 
$D_\perp, \lambda, s_0, z_0, z_p$  are  fixed by the DIS analysis in~\cite{Stoffers:2012zw} 
as listed in Table~\ref{table:1}. For the light vector mesons, we have set $m_{u,d,s}$ at their
constituent values, and $m_{c,b}$ at their PDG values. 
The value of $\kappa$ is adjusted to reproduce the best value for the vector meson
decay constants. The vector masses $M_V$ are then fixed by (\ref{KAPPA}) as listed in 
Table~\ref{table:1}. In our holographic set up, the lower decay constants for the heavier mesons
imply smaller values of $\kappa$ (string tension) for $J/\Psi, \Upsilon$ in comparison to the 
$\rho$ for instance. Since $f_V^2$ is a measure of the compactness of the wavefunction at the origin 
this is reasonable, although the spread in the transverse direction appears to be larger in the absence
of the Coulombic interactions which are important for $J/\Psi, \Upsilon$.
Finally, the string coupling $g_s$ is adjusted to reproduce the overall
normalization of the cross section for each vector meson channel.

\subsection{Radiative widths}

In terms of (\ref{FVL}), the radiative decay width $\Gamma(V\rightarrow e^{+}e^{-})$ is
\be
\label{RATIOGV}
\frac{\Gamma}{e_V^2} 
=\frac{4\pi\alpha^2}{3M_V}\frac{f_V^{L2}}{e_V^2}
\ee
We note that
(\ref{FVL}) is finite in the heavy quark limit as expected from the Isgur-Wise symmetry.
Using (\ref{001}),  (\ref{RATIOGV})  gives
\be
\label{RATIOEXP}
\left(\rho: 9.32;  \omega: 8.30; \phi:  10.6; J:
    3.71;  \Upsilon:  0.51\right) 
\,{\rm KeV}
\ee
The emprical ratios of the width to the squared charge  are

\be
\label{RATIOEXP}
\left(\rho: 13.2; \omega: 12.8; \phi: 11.8; J: 10.5;
    \Upsilon: 10.6\right) 
\,{\rm KeV}
\ee
with $e_V$ fixed by (\ref{EV}). The holographic decay widths are in agreement with the empirical ones
for the light vector mesons $\rho, \omega, \phi$, but substantially smaller for the heavy vector mesons 
$J\Psi, \Upsilon$. This maybe an indication of  the strong Coulomb corrections in the heavy quarkonia 
missing in our current holographic construction. One way to remedy this is through the use of improved
holographic QCD~\cite{KIRITSIS}.


\subsection{$\gamma p\rightarrow \rho p, \omega p$}

\begin{figure}[!htb]
 \includegraphics[width=75mm]{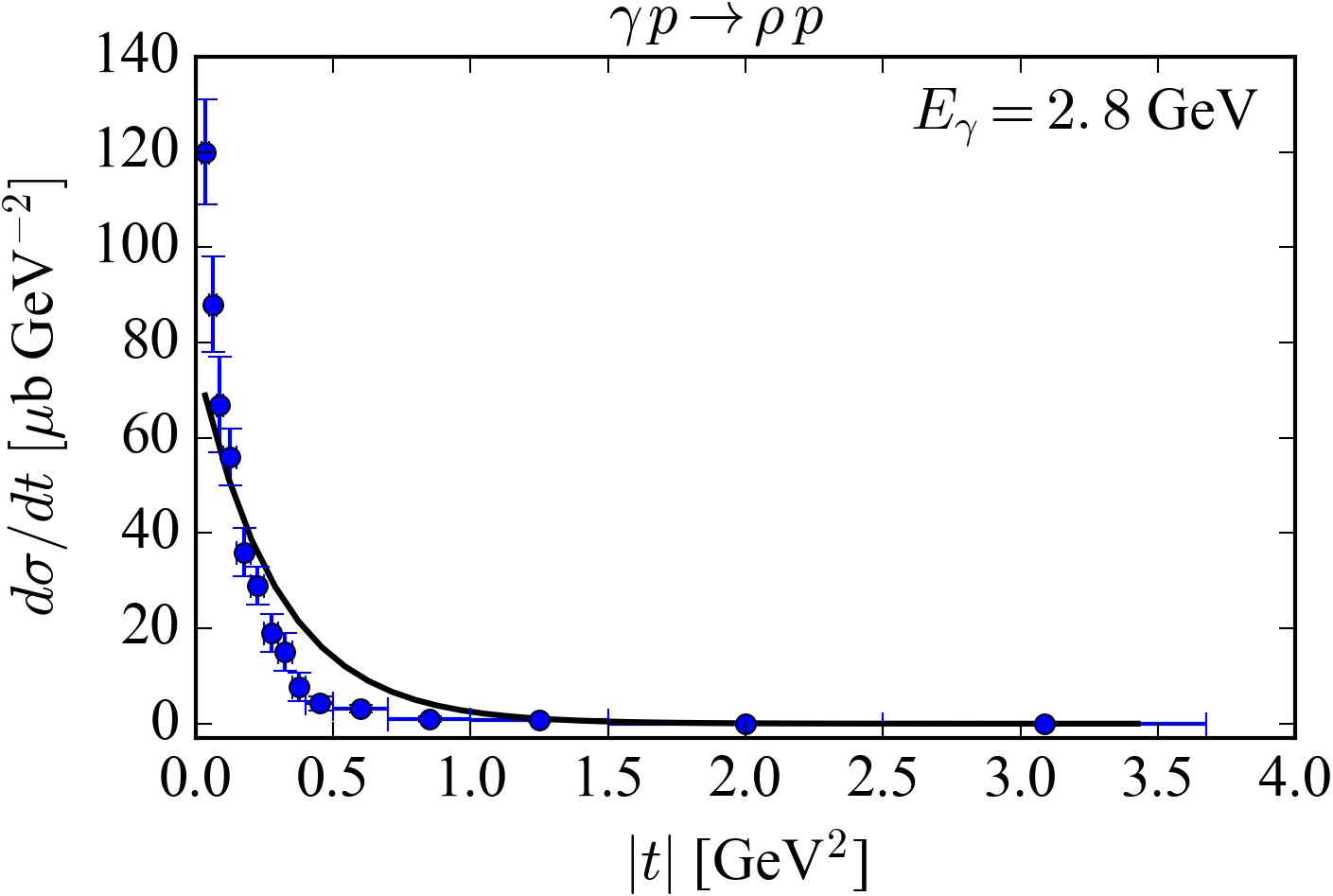}  
  \caption{Differential cross section for $\gamma p \to \rho p$ versus $|t|$
  for $E_\gamma=2.8\,{\rm GeV}$: the solid line is this work, the filled circle are
  the data. Data are taken from \cite{Ballam:1971yd}.}
  \label{fig:3}
\end{figure}

\begin{figure}[!htb]
 \includegraphics[width=75mm]{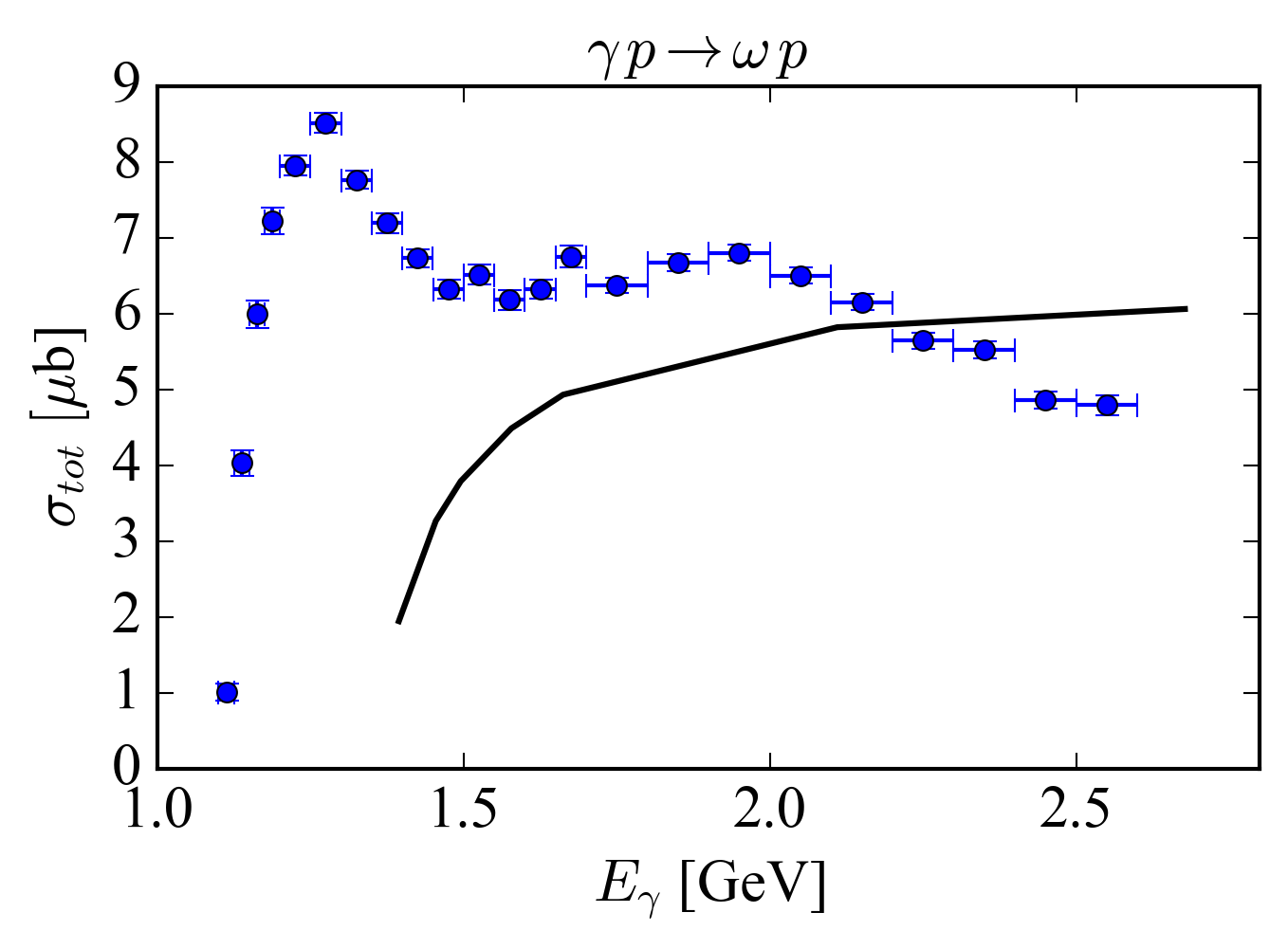} 
  \caption{Total cross section for  $\gamma p \to \omega p$ versus $E_\gamma$. 
The solid line is this work, the filled circles are the data.
Data are taken from \cite{Barth:2003kv}.}
\label{fig:4}
\end{figure}

In Fig.~\ref{fig:3} we show the differential $\rho$-photoproduction versus $|t|$ for $E_\gamma=2.8$ GeV. 
At this energy the photon size is of the order of the hadronic sizes and sensitive to non-perturbative physics. 
In Fig.~\ref{fig:4} we show the total cross section for $\omega$-photoproduction in the range of low mass
photons. The discrepancy close to treshold maybe due to  t-channel sigma-exchange and the s-channel
photo-excitation of the $\Delta(1230), N(1520), N(1720)$ in the intermediate nucleon state, not retained in our analysis.
Note that both the $\rho$ and $\omega$ have comparable transverse sizes with 
$1/\kappa\approx \frac 13\,{\rm fm}$ but very different decay constants. We expect their differential and total cross
sections to be in the ratio of their decay constant, say $f_\omega^2/f_\rho^2\approx \frac 1{10}$.

\subsection{$\gamma p\rightarrow \phi p$}

\begin{figure}[!htb]
 \includegraphics[width=78mm]{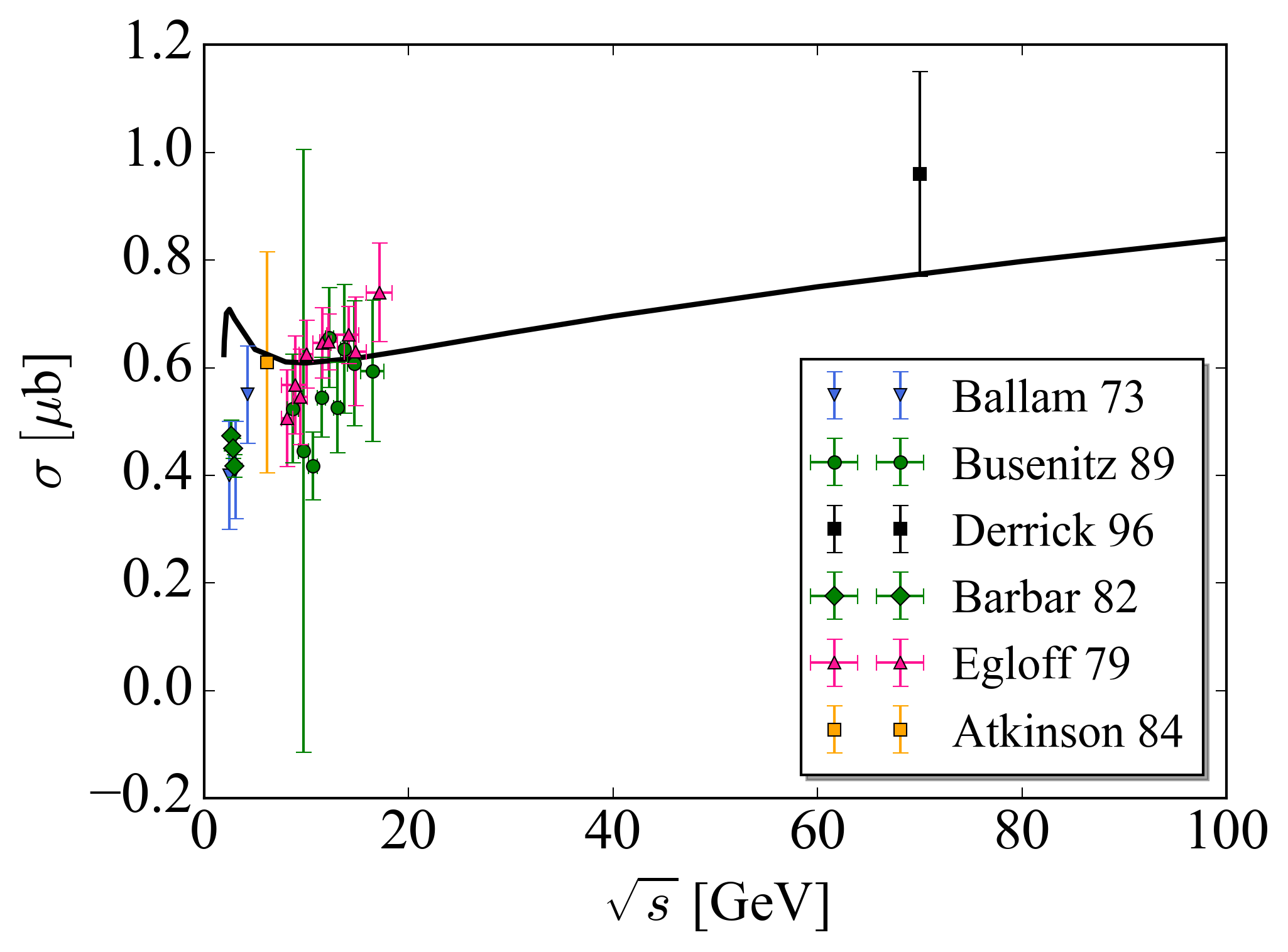}
  \caption{Total cross section for $\gamma p \to \phi p$ from
    threshold to $\sqrt{s}=100$ GeV.
Data are taken from \cite{Ballam:1972eq, Derrick:1996af, Busenitz:1989gq}. 
}\label{fig:7}
\end{figure}

\begin{figure}[!htb]
 \includegraphics[width=75mm]{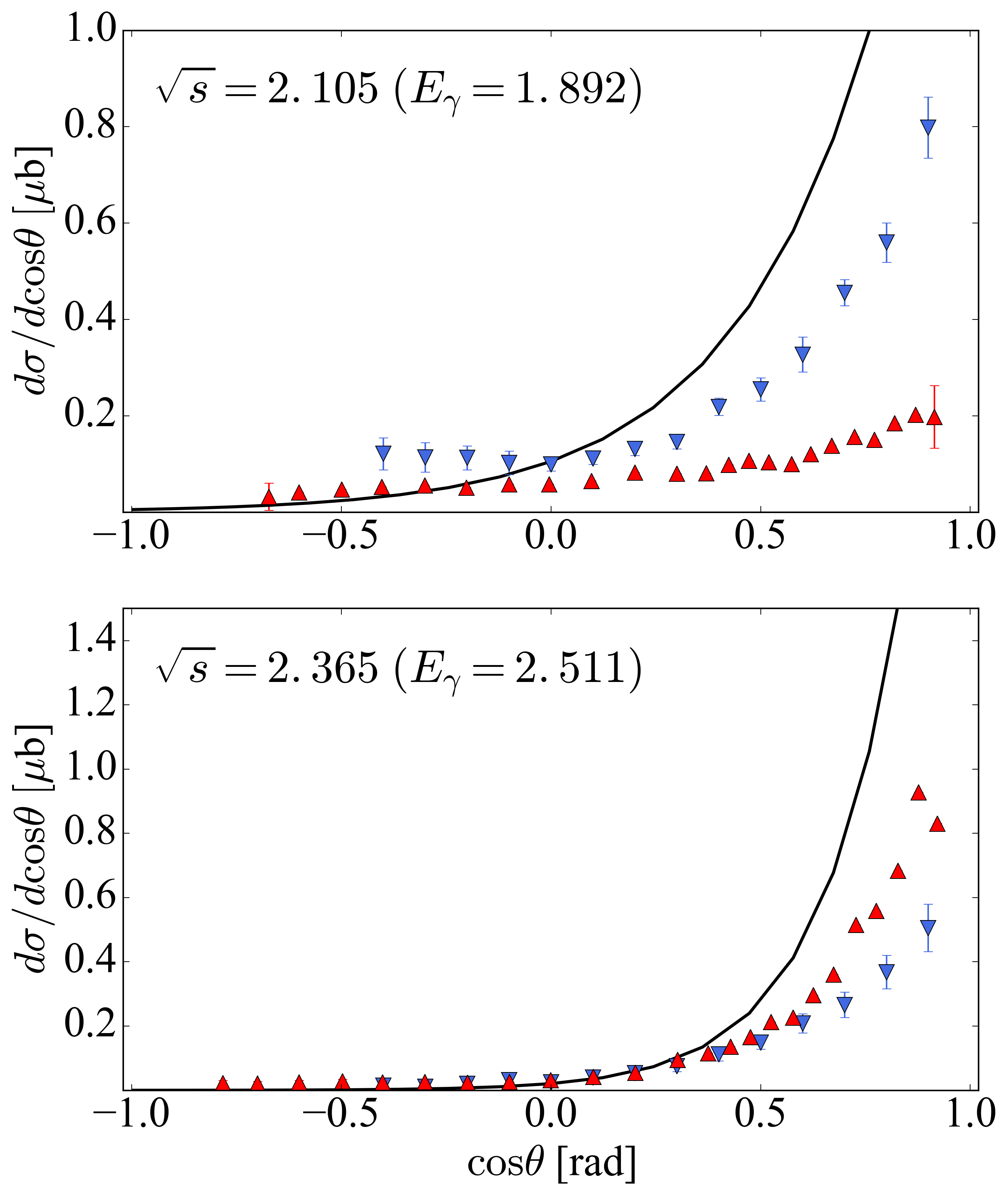}
  \caption{Differential cross section for $\gamma p \to \phi p$ in low
    energy region.
The black solid line is the present work.
Data at $\sqrt{s}=2.015$ and 2.365 GeV (red up triangle) are taken
from the charged mode \cite{Dey:2014tfa}  and at $\sqrt{s}=2.13$ and 2.38 GeV (blue down
triangle)
from the neutral mode \cite{Seraydaryan:2013ija} in the CLAS Collaboration.
}\label{fig:5}
\end{figure}

\begin{figure}[!htb]
 \includegraphics[width=75mm]{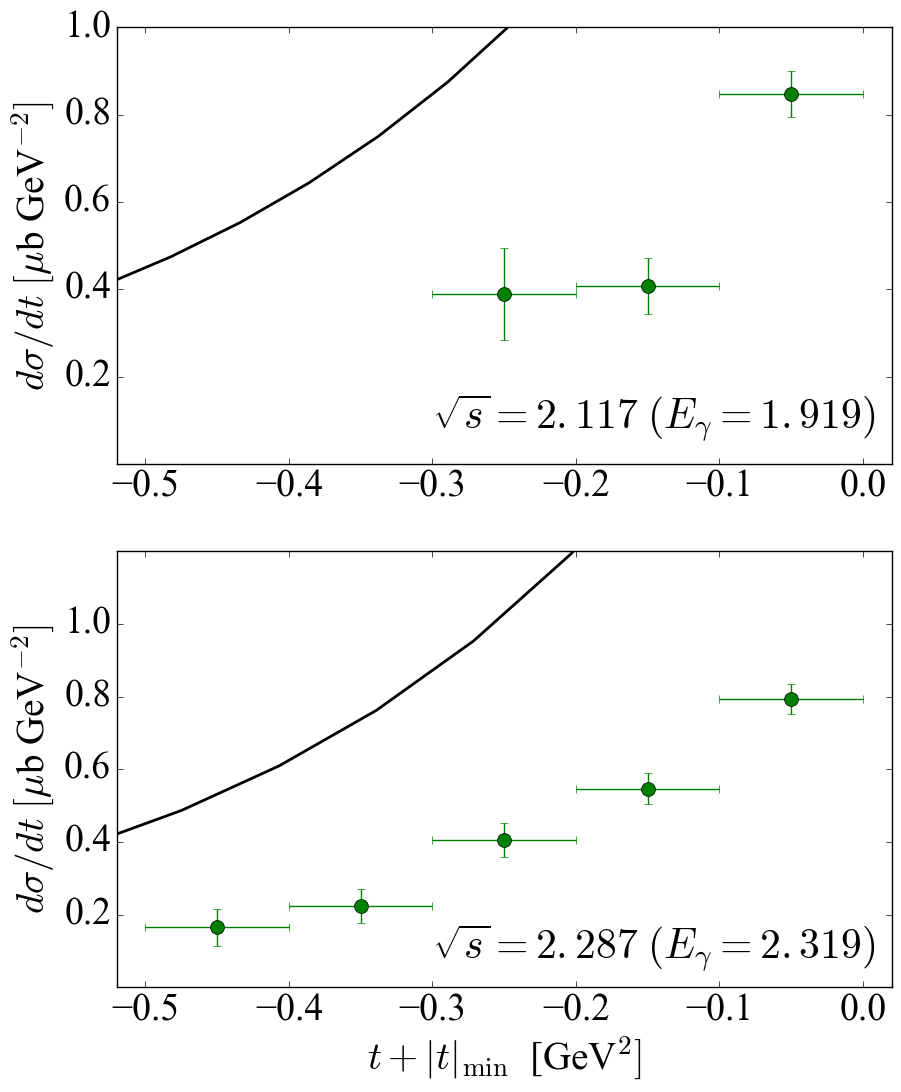}  \caption{Differential cross section for $\gamma p \to \phi p$ in low
    energy region.
The black solid line is the present work.
The data is from \cite{Mibe:2005er}.
The units for the photon energy and $\sqrt{s}$ are GeV.
}\label{fig:6}
\end{figure}

\begin{figure}[!htb]
 \includegraphics[width=75mm]{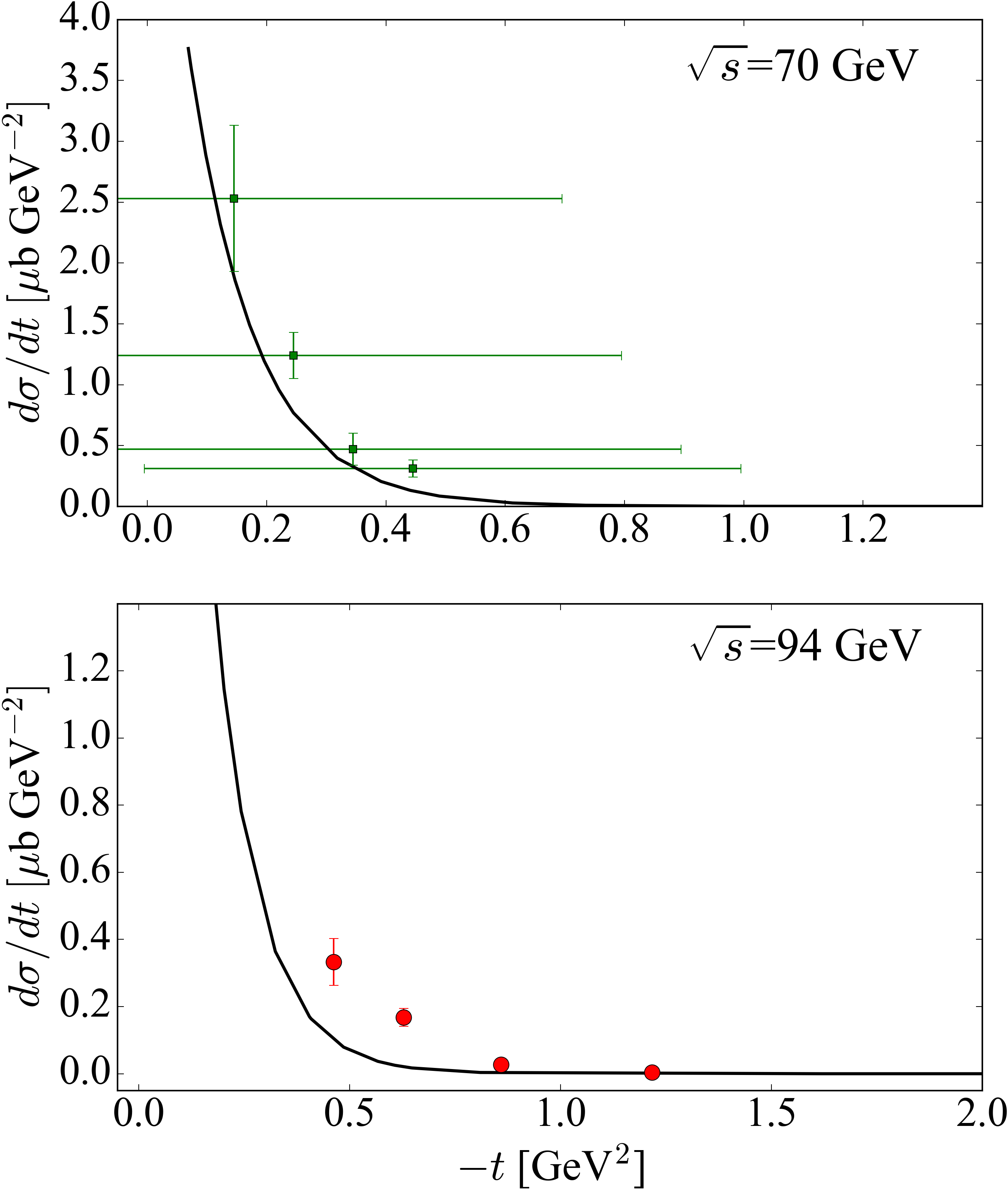}
  \caption{Differential cross section for $\gamma p \to \phi p$ in the
    high energy region.
Data are taken from \cite{Derrick:1996af,Breitweg:1999jy}
}\label{fig:8}
\end{figure}

In Figs. \ref{fig:7}-\ref{fig:8}, we present the total and
differential cross section for the $\phi$-photoproduction $\gamma p \to \phi p$ process.
In Fig. \ref{fig:5} we compare our results to the available  CLAS  and  LEPS
data. Our results agree with the backward angle data well, but overshoot the forward angle data.
In Fig.~\ref{fig:6}-\ref{fig:8}, the differential cross sections are shown. The agreement at large $\sqrt{s}$ 
probes mostly the Pomeron exchange. Note that 
our overall fit to the $\phi$-decay constant implies
a transverse size for the $\phi$ that is comparable to the $\rho, \omega$ 
sizes, which is reasonable. The differential and total cross sections are expected to be in the ratio
of the squared decay constants $f_\phi^2/f_\rho^2\approx \frac 15$ or $f_\phi^2/f_\omega^2\approx \frac 12$.

\begin{figure}[!htb]
 \includegraphics[width=75mm]{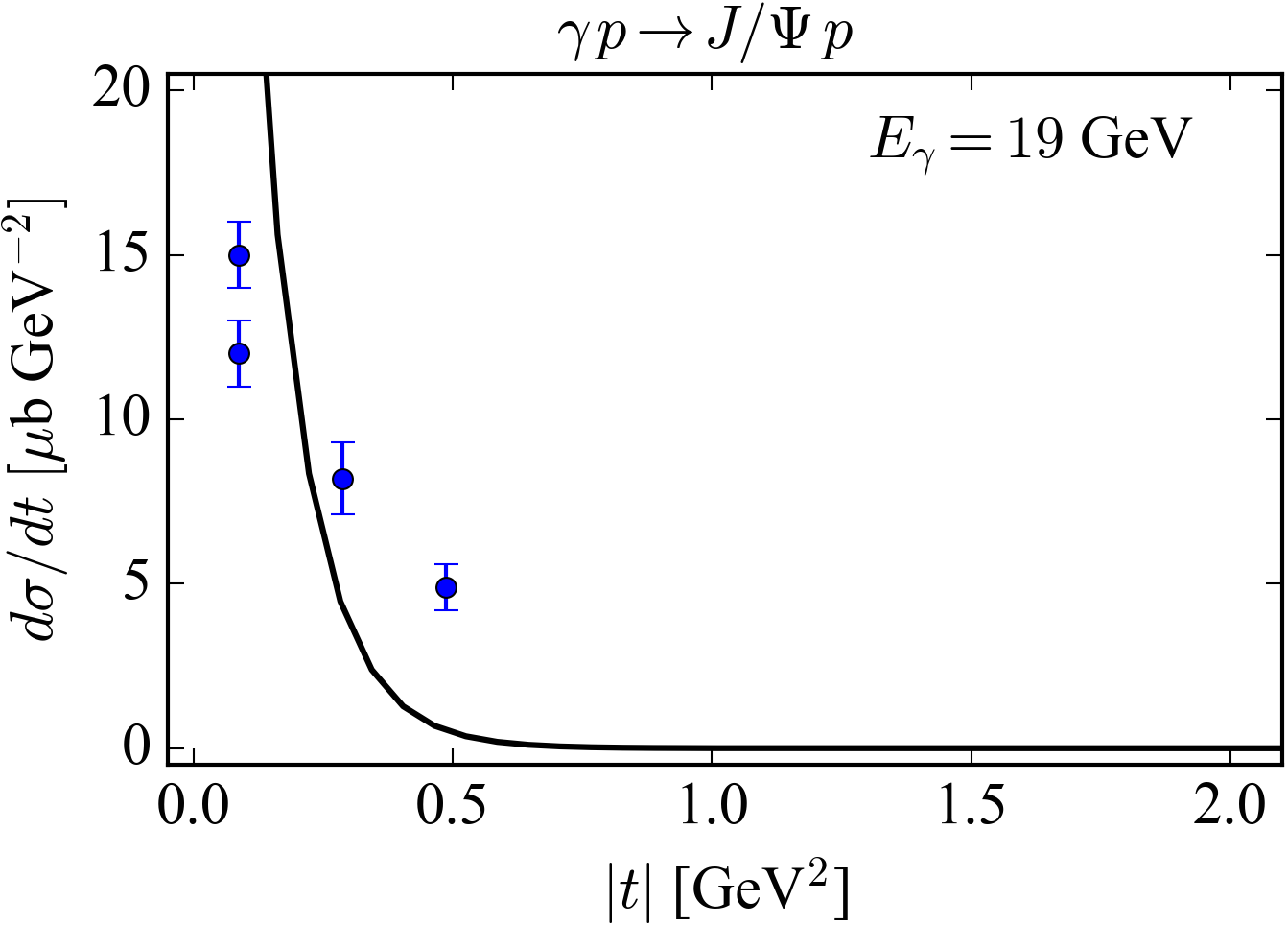} 
  \caption{Differential cross section for $\gamma p \to J/\Psi p$.
The data is from \cite{Camerini:1975cy}.
}\label{fig:9}
\end{figure}

\subsection{$\gamma p\rightarrow J/\Psi p, \Upsilon p$}

\begin{figure}[!htb]
 \includegraphics[width=75mm]{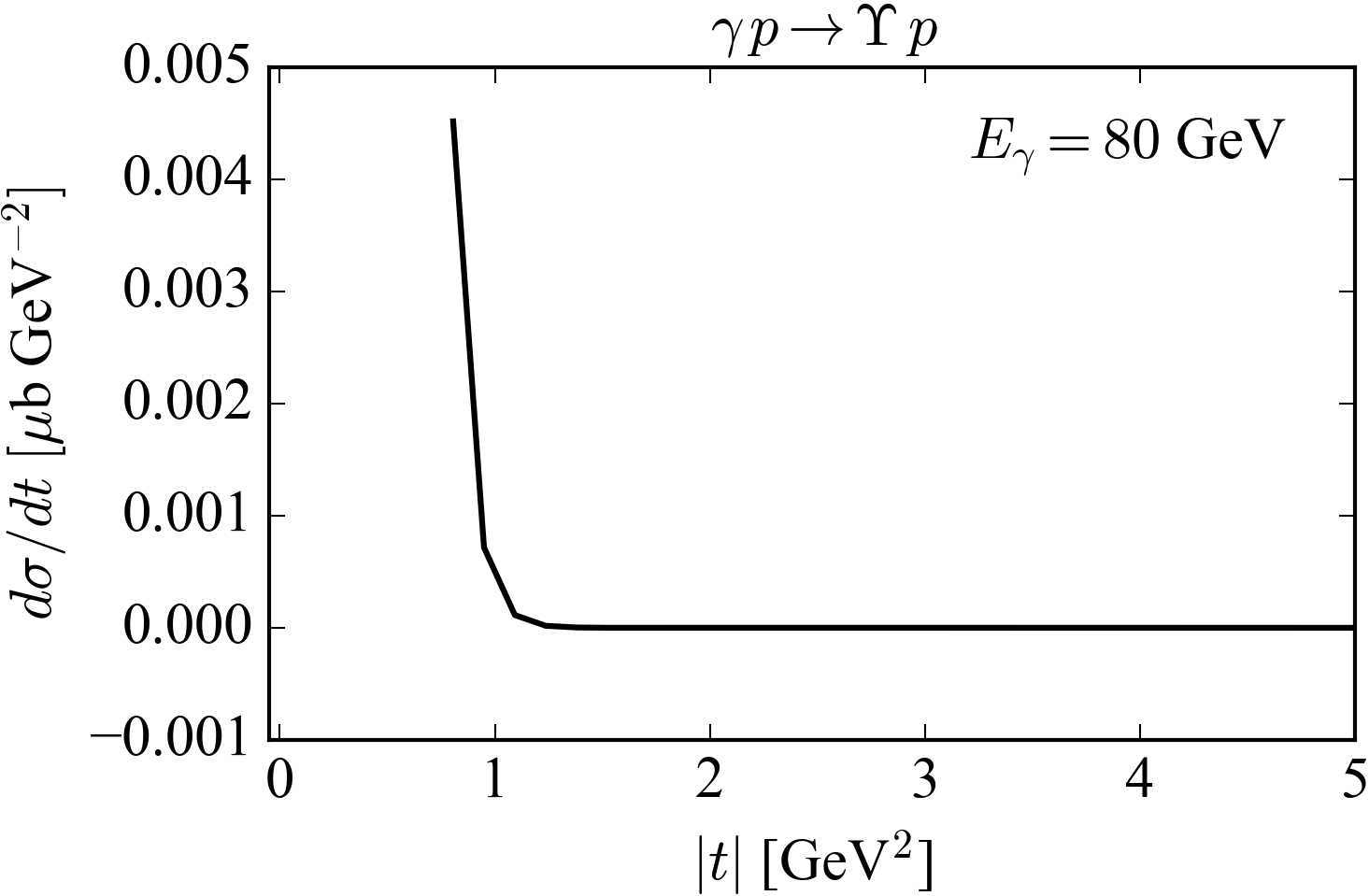}
  \caption{Differential cross section for $\gamma p \to \Upsilon p$.
The black solid line is the present work.
}\label{fig:10}
\end{figure}

In  Figure \ref{fig:9} we show the differential corss section for 
$\gamma p \to J/\Psi p$ process, and in Figure \ref{fig:10} we show the differential corss section for 
$\gamma p \to \Upsilon p$ process. We note that $2m_f=2.58, 8.83\,$GeV respectively, so 
$\sqrt{s}>10\,$GeV are necessary to eikonalized the heavy quarks. These results are only exploratory,
since the transverse sizes of the $J/\Psi, \Upsilon$ are large in our current construction as we noted earlier.
To remedy this shortcoming requires including the effects of the colored Coulomb interaction which is important
in these quarkonia states. In holography this can be achieved through the use of  improved holographic QCD~\cite{KIRITSIS}
which is beyond the scope of our current analysis.

\section{Conclusions}

In QCD the diffractive photo-production of vector mesons  on protons at large $\sqrt{s}$ is described as the
scattring of two fixed size dipoles running on the light cone and exchanging a soft pomeron. In a given hadron
the distribution of fixed size dipoles is  given by the intrinsic dipole distribution in the light cone wavefunction. 
The soft pomeron exchange and the intrinsic dipole distribution are non-perturbative in nature. We use the 
holographic construct in AdS$_5$  to describe both. 

The soft Pomeron parameters  used in this work were previously  constrained by the DIS data~\cite{Stoffers:2012zw},
so the extension to the photoproduction mechanism is a further test of the holographic construction. 
The new parameter characterizing the transverse size of the vector mesons was adjusted to reproduce the meson 
radiative decays and found to be consistent with the expected string tension characteristic of the vector Regge trajectory.
Comparison of our results to the data for photoproduction of vectors show fair agreement with data for the
$\rho, \omega, \phi$, although the inclusion of Reggeon exchanges 
may improve our description at low photon masses near treshold. At high photon masses, perturbative QCD scaling laws are expected. 
Our analysis of the photoproduction of $J/\Psi, \Upsilon$ is limited since the present construction does not account
for the substantial Coulomb effects for these quarkonia. We hope to address this issue and others next.

\section{Acknowledgements}

This work was supported by the U.S. Department of Energy under Contract No.
DE-FG-88ER40388.
HYRyu and CHL were partially supported by the National Research Foundation of Korea (NRF) grant funded by the 
Korea government (MSIP) (No. 2015R1A2A2A01004238 and
No. 2016R1A5A1013277). \\

 \vfil

\end{document}